\documentclass[10pt,twocolumn]{article}
\usepackage{graphicx}
\usepackage{geometry} 
\usepackage{color}
\usepackage[square, comma, sort&compress]{natbib}
\pdfoutput=1
\title{Were people imitating others or exercising rational choice in on-line searches for `swine flu'?}
\author{R. Alexander Bentley\\Anthropology Department \\ Durham University, UK\\r.a.bentley@durham.ac.uk \and Paul Ormerod\\Volterra Consulting Ltd.\\ London, UK\\pormerod@volterra.co.uk}
\begin{document}
\maketitle
\textbf{Two general patterns have been identified for the adoption and subsequent abandonment of ideas or products within a population.   One is symmetric, so that concepts or products which are adopted rapidly then decline rapidly from their peak, and those which are slower to move to their peak decline more slowly.  The other is asymmetric, where the decline from the peak is considerably slower than is the rise to the peak, and vice versa. We posit that these contrasting patterns arise from two fundamentally different modes of behaviour which are used by humans in making choices in different contexts.  Namely, choice based on the imitation of the choices of others versus purposeful selection based upon the inherent attributes of the concept or product.  We illustrate the proposition with the example of internet searches for the phrase `swine flu' in a wide range of countries across the world.  The methodology offers a general heuristic for distinguishing between these two general and contrasting modes of behavioural choice.}

\section{Introduction}
Qualitatively, two general patterns have been identified for the uptake and subsequent decline of cultural products and phenomena.   One is symmetric, so that concepts or products which are adopted rapidly then decline rapidly from their peak, and those which are slower to move to their peak decline more slowly.  Berger and Le Mens \cite{BergerLeMens_2009} illustrate this with the example of first names in the US during the 20th century, and Bentley and Ormerod \cite{BentleyOrmerod_2009} give examples of hill goers completing lists of mountains over a specific height.  

But the other is asymmetric, where the decline from the peak is considerably slower than is the rise to the peak, and vice versa.  Sornette \textit{et. al.} \cite{Sornette_etal_2004} and Deschartres and Sornette \cite{DeschartresSornette_2005} give examples of the two general patterns in on-line book sales, and offer an explanation based upon a distinction between exogenous and endogenous shocks to behaviour.

Berger and Le Mens  \cite{BergerLeMens_2009} account for their findings in terms of psychological effects in the population: essentially, if people perceive something rising quickly, they will tend to discard it quickly.  Conversely, if it has risen in popularity slowly, it will be seen as more substantial and lasting and thus remain popular longer.

Here we describe a fundamental dichotomy in human behaviour in making choices, namely choice based on the imitation of the choices of others versus purposeful selection based upon the inherent attributes of the concept or product.  

Both these behavioural models have extensive support in different contexts in the scientific literature, particularly that of anthropology and economics.  We posit that agents adopt one or other of these modes of choice in different contexts.  Of course, in practise there will not often be a pure `either/or' dimension to the model of behavioural choice, and agents may use elements of both.

However, positing such an `either/or' decision generates a straightforward way of identifying in consumer choice the relative importance of imitation and deliberate selection, based upon the relative speeds of take up to peak and subsequent decline from the peak of a concept or product.  In other words, which of the two stylised patterns described above is observed will depend upon the relative importance attached by agents to these two basic methods of choice.

We illustrate the behavioural model with data on the pattern of Google searches for `swine flu' across a range of countries across the globe. 
 
 \section{Imitation versus deliberate selection}
 
 A parsimonious behavioural model of choice has been shown to be able to account for a wide range of cultural phenomena, evolving over massively different time spans.  Referred to by some anthropologists as `cultural drift', we label this model `imitation', since this is the main driver of choice under this mode of behaviour.

Essentially, in making a choice using the `imitation' mode of behaviour, the agent samples from the population and adopts the choice already adopted by the agent which he or she selects at random.  But in addition, a small percentage of the time, the agent will adopt a unique invention.  Evidence from a range of disciplines suggest that this frequency is of the order of 5 per cent or even less \cite{Rogers_1962, SrinivasanMason_1986, Eerkens_2000, Diederen_etal_2003}.

This evolutionary model has been very useful both for archaeologists measuring cultural change over centuries \cite{Neiman_1995, Lipo_et al_1997, ShennanWilkinson_2001, BentleyShennan_2003}, and in studies of modern data sets over scales ranging from decades down to weeks \cite{HahnBentley_2003, Bentley_et al_2004, Bentley_et al_2007, BentleyOrmerod_2009, Herzog_etal_2004}. 

Under this model of choice, the more prevalent something is in the population, the longer it takes to disappear; that is, while it is still `out there' to be imitated.  So all else being equal (including peak height), those things that take longer to go up (in terms of additions per time unit) will generate larger cumulative frequencies in the population, and the therefore take longer to decline.  In general, when choice is mainly driven by `imitation', the pattern of adoption to and decline from the peak of a cultural phenomenon should tend to be symmetric.

In contrast to this `imitation' model, humans in other contexts adopt a `selective' mode of behaviour, which is cognate with the economists' concept of rationality: agents make deliberate choices as individuals depending upon the perceived attributes of the concept of product being considered. 

Under selection, the speed of the rise in popularity should reflect the perceived significance of the idea. So, popularity which rises quickly due to the perceived importance/significance will in general decline more slowly as that significance fades.   Conversely, things that go up more slowly, being less urgent or important, should tend to decline relatively faster.  

These two contrasting modes of behaviour in the exercise of choice, imitation and deliberate selection, can be illustrated by the rise and fall in popularity of first names in America in the 20th century.  

As Berger and Le Mens  \cite{BergerLeMens_2009} point out, the name `Charlene' took two decades to rise to its peak U.S. popularity around 1950, and about two more decades to decline to a similar level, whereas `Tricia' rose and fell from its peak all in less than two decades.  We postulate that the speed of adoption reflects the speed of abandonment because ÔCharlene' generated a larger cumulative popularity over that much longer time span than `Tricia'. This is why it took proportionately longer to decline (the 1990 US Census shows about 3.2 times more Charlenes than Tricias in the general population). 

The adoption-decline pattern for these two popular names, which spread by imitation, contrast markedly with names that enjoyed brief bursts of popularity due to deliberate selection.  Figure 1 shows the U.S. popularity of some previously-rare first names that were clearly affected by selection during the 20th century, either due to the election of a president (Woodrow Wilson in 1912, FDR in 1932), military success (Dwight Eisenhower in the 1940s, whose name was boosted again by his presidency), and music celebrity (Elvis Presley in the 1950s, Britney Spears in the late 1990s).  In each case, the trajectory is asymmetrical Ð a rapid rise to popularity, followed by a slow decline.\\

 FIGURE 1 HERE\\

 \section{Swine flu searches on Google}

The new, very powerful (although perhaps not the most precise) online analysis tool offered by Google, called ``Google Trends,'' provides a calendar timeline of search results for any search term, which is sub-divisible by country (or even regions within each country).  Google Trends does not provide the exact numeric data on the volume of searches.  It provides a relative account of search volume, which is plotted on a scale that appears to normalise the results such that the maximum search volume = 1 unit.  

In a cross-country comparison of course, this is exactly how we would want the data displayed because it is not necessarily the number of people in a country who are doing searches that matters (lest China and the US come out on top of every list), but the relative proportion of the peak popularity at any given time. 

As an example, we examined the recent explosive spread of interest in `swine flu' towards the end of April 2009, reverberating into the first week of May. We used data from 30 countries, a sufficient number to form a reasonable sample size.  These span the globe, from North, Central and South America, Western Europe, Middle East, South and East Asia and Australasia.   Some countries we would have liked to include in the analysis such as Russia had insufficient searches over whole or part of the relevant period for the data to be made available.  The countries used are set out in Table 1 below.

On this basis, we can very quickly assess the popularity rise and fall of searches for a certain idea, such as ``swine flu,'' as shown In Figure 2.\\

FIGURE 2 HERE\\

    As shown in Figure 2, the rise and fall of searches on `swine flu' can be characterised by four key landmarks, including its initial appearance at a detectable level (1), the date when it first reaches half (2) of its peak popularity (3), the date when it has lost half of that peak popularity (4).  A final point of interest is to take a set date some time after the decline (8 May in the example above), and estimate how much activity remains, as a fraction of peak activity.  

The convenience of this method is that it does not matter what the actual numbers of searches are on the y-axis, we just make use of what limited data that Google freely provides, which are dates and the relative volume of searches.  Because adoption and decline curves are generally non-linear, with change roughly in proportion to the current level, the `half-life' dates should be roughly comparable between countries Ð whether they have huge number of internet users or relatively few.

 Generally speaking, all the curves from different countries look like that in Figure 2, but with subtle differences in the date of each of the four landmarks, and in the level of `resonating' activity on 8 May.

\begin{table*}
\caption{Date of the landmarks, from 1 April (e.g., 1 May = day 31), for ``Swine flu'' search time-series, along with the remaining search volume on 8 May relative to peak height.}
\begin{tabular}{@{\vrule height 8pt depth5pt  width0pt}lrcccc}
\\\hline
Country & (1) first appearance & (2) half-up & (3) peak & (4) half down & (5) 8 May (fract. peak) \\\hline Canada & 24 & 26.0 & 30 & 31.5 & 13.0\% \\ UK & 24 & 26.5 & 30 & 32.0 & 19.2\% \\ US & 24 & 26.5 & 30 & 31.5 & 8.2\% \\Mexico & 24 & 25.5 & 28 & 31.5 & 11.6\% \\ Australia & 25 & 26.5 & 28 & 32.0 & 8.8\% \\ France & 25 & 26.5 & 30 & 31.5 & 13.1\% \\ Germany & 25 & 26.5 & 30 & 31.5 & 16.3\% \\ Japan & 25 & 26.5 & 30 & 31.5 & 17.6\% \\ Netherlands & 25 & 26.5 & 30 & 31.0 & 9.1\% \\ Singapore & 25 & 27.0 & 30 & 33.0 & 9.4\% \\ Spain & 25 & 26.0 & 30 & 32.0 & 14.9\% \\ New Zealand & 25 & 25.5 & 28 & 30.5 & 6.7\% \\ Belgium & 26 & 26.0 & 30 & 31.0 & 0.0\% \\ Brazil & 26 & 26.5 & 30 & 32.0 & 39.4\% \\ China & 26 & 26.5 & 29 & 33.0 & 15.6\% \\ Italy & 26 & 26.0 & 27 & 32.0 & 0.0\% \\ South Korea & 26 & 26.5 & 29 & 31.5 & 0.0\% \\ India & 26 & 26.5 & 32 & 34.5 & 22.0\% \\ UAE & 26 & 26.5 & 28 & 33.5 & 18.2\% \\ Egypt & 26 & 28.0 & 33 & 35.0 & 30.5\% \\ Israel & 26 & 26.0 & 28 & 30.5 & 0.0\% \\ South Africa & 26 & 27.0 & 30 & 31.0 & 12.3\% \\ Philippines & 26 & 26.5 & 30 & 32.5 & 30.7\% \\ Malaysia & 26 & 26.5 & 29 & 32.5 & 15.0\% \\ Thailand & 26 & 26.0 & 30 & 33.0 & 0.0\% \\ Finland & 26 & 26.0 & 30 & 32.0 & 0.0\% \\ Taiwan & 27 & 27.0 & 28 & 32.0 & 0.0\% \\ Indonesia & 27 & 27.0 & 28 & 32.0 & 0.0\% \\ Saudi Arabia & 27 & 27.0 & 28 & 33.5 & 0.0\% \\ Turkey & 27 & 27.0 & 30 & 32 & 0.0\% \\
\end{tabular}
\end{table*}

The data can be downloaded on a daily basis, which requires approximations to be made when selecting the timings of the two `half peak' values described above.  Essentially, when the level of search on a given day was very similar to half that on the peak day, we assigned this day to be a `half peak', whether up or down.  In cases where the half peak value was clearly spanned by two successive dates, 26 and 27 April say, we assigned the `date' of `26.5 April' to the relevant half peak.  More sophisticated interpolation could certainly be used, but it must be remembered that this is not data generated by a controlled scientific experiment, and so should not be assigned undue precision. We set out the data in Table 1.

The numbers in Table 1 all look fairly similar, but interesting patterns come from sorting these data based on the different columns, and noting patterns in the countries at either end of the list.  

\textbf{Origins}: As indicated by landmark 1, the searches for `swine flu' first became significant in the developed, English-speaking countries of US, UK and Canada (24 April), and was last to register in non-English speaking and/or less developed countries (26/27 April). 

\textbf{Speed of adoption}: Possibly as a result of being the first to buzz about it,  the countries that took the most days to go from first appearance to half-peak (landmark 2) were the UK and the US, both needing about 2.5 days to reach half-peak. The countries that most quickly reached half-peak, in less than a day, were Belgium, Italy, Israel, Thailand and Taiwan.  

\textbf{Speed of decline}: As defined by the time between Landmark 3 (half-down) and the peak, the swine flu searches resonated longer in Asian countries, including Thailand, Singapore, Malaysia, China, the Philippines, and Japan, whereas the buzz had died down most thoroughly in European-American countries. 

Figure 3 shows a clear negative correlation across countries in the time taken for search values to decline from peak to half-peak level and the time taken to move from half peak level on the way up to the peak itself.\\

FIGURE 3 HERE\\

The simple correlation is $-0.74$, and the $r^2$ value in the non-linear regression is $0.70$ with an equation standard error of $0.751$ and an effective number of parameters of $2.4$.  Analysis of variance shows that the null hypothesis of a linear relationship is rejected at a $p$-value of $0.002$.
  
\section{Conclusion}
The asymmetrical relationship between the speed of the rise to the peak search level and the speed of the subsequent decline indicates that the dominant mode of behaviour motivating these `swine flu' internet searches was that of deliberate selection.  Agents carrying out the searches were primarily doing so to discover information about the `product' \textit{i.e.} swine flu, rather than imitating the actions of others.

The methodology we have proposed here offers a heuristic for distinguishing between these two general and contrasting modes of behavioural choice. It could be applied more generally to data on cultural concepts and products to distinguish `markets' in which agents are primarily driven by `fashion', in other words imitating the actions of others, and `markets' in which the main motivation of agents in their choices is deliberate selection according to the attributes of the concept/product being chosen.

 \begin{figure*}
  \begin{center}
  \includegraphics[width=6in]{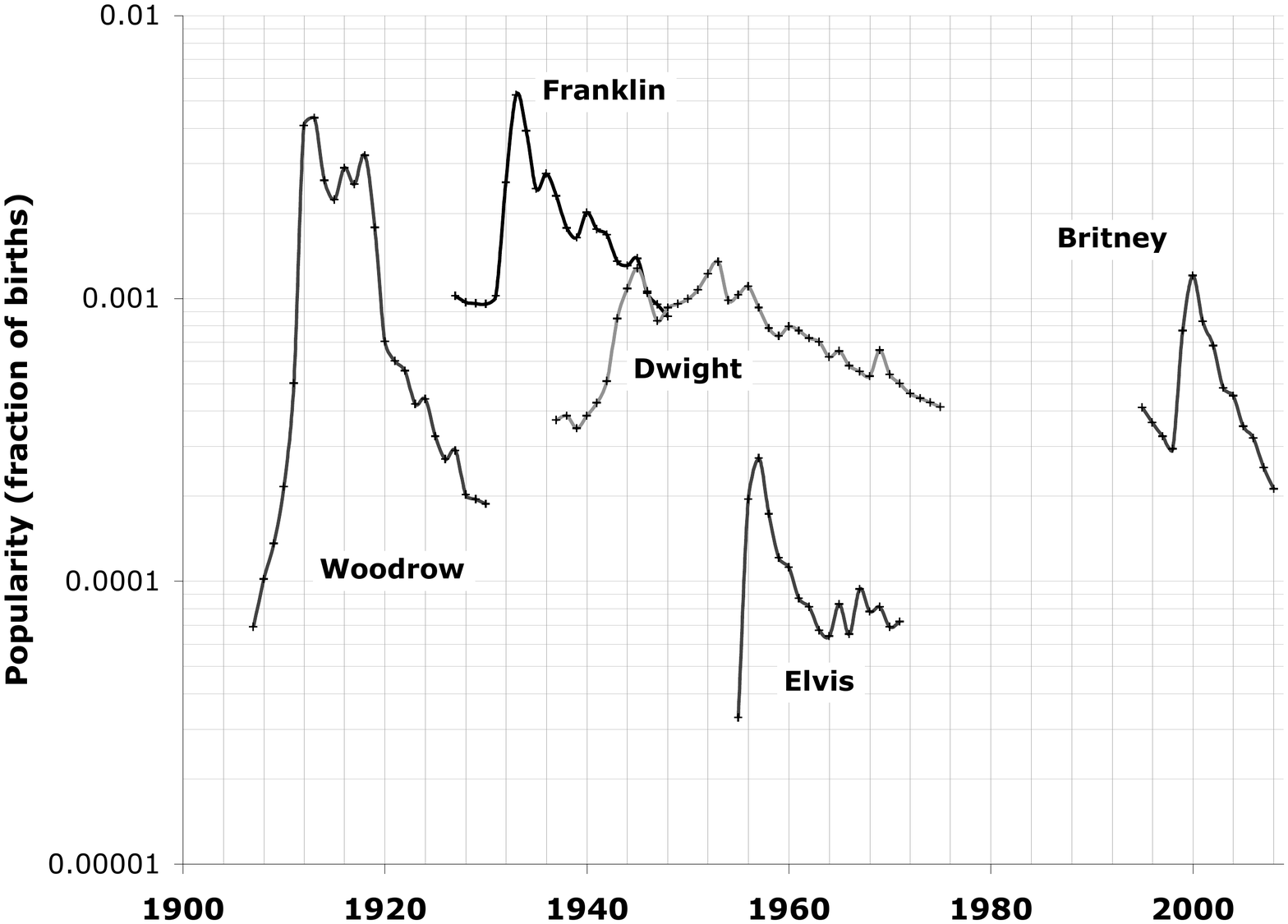}
  \end{center}
  \caption{Popularity (in the U.S.) of some first names that have clearly been affected by selection, due to the fairly abrupt appearance of a famous person with that name. The asymmetry of these trajectories contrasts with the symmetry of the majority of names \cite {BergerLeMens_2009} affected mainly by drift \cite{HahnBentley_2003}.}
  \end{figure*}

\begin{figure*}
  \begin{center}
\includegraphics[width=6in]{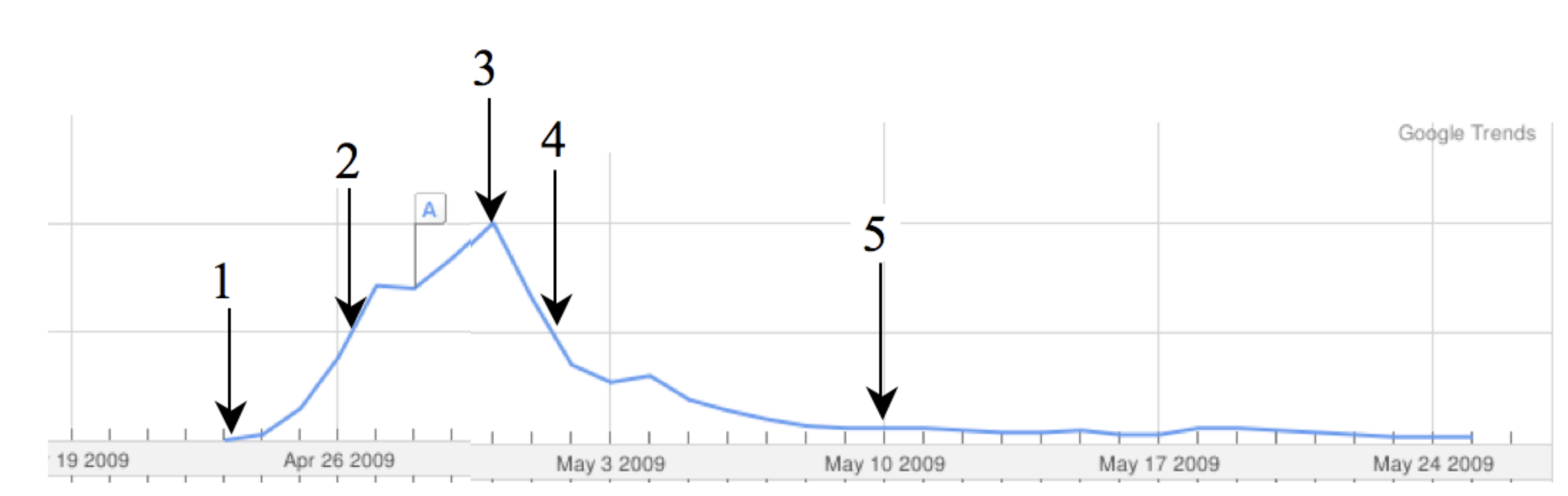}
  \end{center}
 \caption{The Google Trends result for `Swine flu' during April-May 2009. This plot shows the aggregated world results for all countries--individual country results are subtly different, and together they provide the range of uptake/decline rates for this study.}
 \end{figure*}

\begin{figure*}
 \begin{center}
 \includegraphics[width=5in]{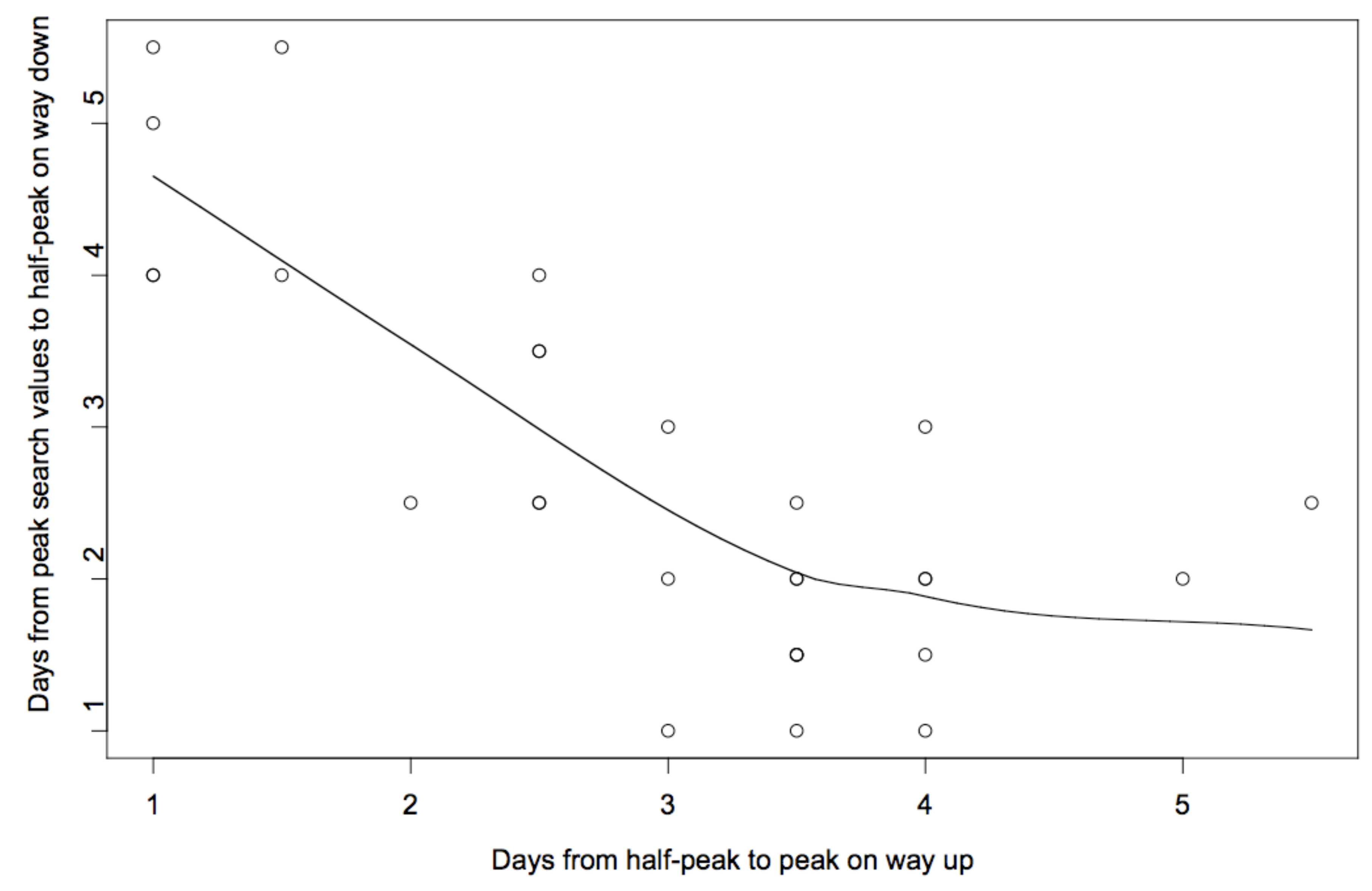}
 \end{center}
 \caption{`Swine flu' Internet searches during the outbreak of April-May 2009, as recorded by Google Trends. Each datum represents a different country, with the x-axis showing the number of days between landmarks 2 and 3 (see Figure 1) and the y-axis showing days between landmarks 3 and 4.  The solid line is the optimal non-linear fit using the technique of local linear regression in the statistical package S-Plus.}
  \end{figure*}


\begin{thebibliography}{}

\bibitem{BergerLeMens_2009} Berger, J., G. Le Mens (2009). How adoption speed affects the abandonment of cultural tastes. {\it PNAS} 106: 8146-8150. 

\bibitem{Sornette_etal_2004} Sornette, D., F. Deschatres, T. Gilbert and Y. Ageon (2004). Endogenous versus exogenous shocks in complex networks.  {\it Phys. Rev. Lett.} 93 (22): 228701.

\bibitem{DeschartresSornette_2005} Deschatres, F. and D. Sornette (2005). The dynamics of book sales endogenous versus exogenous shocks in complex networks {\it Phys. Rev. E} 72: 016112.

\bibitem{BentleyOrmerod_2009} Bentley, R.A. and P. Ormerod (2009). Tradition and fashion in consumer choice. {\it Scottish Journal of Political Economy} 56: 371-381.

\bibitem{Rogers_1962} Rogers, E.M. (1962). {\it Diffusion of Innovations} (Free Press, New York).

\bibitem{SrinivasanMason_1986} Srinivasan V. and C.H. Mason (1986). Nonlinear least squares estimation of new product diffusion models. {\it Marketing Science} 5: 169-178.

\bibitem{Eerkens_2000} Eerkens J.W. (2000).  Practice makes within 5\% of perfect. {\it Current Anthropology} 41: 663-668.

\bibitem{Diederen_etal_2003} Diederen P., H. van Meijl, A. Wolters and K. Bijak (2003). Innovation adoption in agriculture. {\it Cahiers d'\'{e}conomie et Sociologie Rurales} 67: 30-50.

\bibitem{Neiman_1995} Neiman F.D. (1995). Stylistic variation in evolutionary perspective. {\it American Antiquity} 60: 7Ð36.

\bibitem{Lipo_et al_1997} Lipo, C.P., M.E. Madsen, R.C.  Dunnell and T. Hunt (1997) Population structure, cultural transmission, and frequency seriation. {\it Journal of Anthropological Archaeology} 16: 301--333.

\bibitem{ShennanWilkinson_2001} Shennan, S.J. and J.R. Wilkinson (2001). Ceramic style change and neutral evolution: A case study from Neolithic Europe. {\it American Antiquity} 66: 577-594.

\bibitem{BentleyShennan_2003} R.A. Bentley and S.J. Shennan (2003). Cultural transmission and stochastic network growth. {\it American Antiquity} 68:459--485.

\bibitem{HahnBentley_2003} M.W. Hahn and R.A. Bentley (2003). Drift as a mechanism for cultural change: an example from baby names. {\it Proceedings of the Royal Society B} 270:S1--S4.

\bibitem{Bentley_et al_2004} R.A. Bentley, M.W. Hahn and S.J. Shennan (2004). Random drift and culture change. {\it Proceedings of the Royal Society B} 271:1443--1450.

\bibitem{Bentley_et al_2007} Bentley RA, Lipo CP, Herzog HA, Hahn MW (2007)  Regular rates of popular culture change reflect random copying. {\it Evolution and Human Behaviour} 28: 151-158.

\bibitem{Herzog_etal_2004} Herzog, H.A., R.A. Bentley and M. W. Hahn (2004). Random drift and large shifts in popularity of dog breeds. {\it Proceedings of the Royal Society B} 271: S353-S356.

\end{thebibliography}
\end{document}